%% file: main.tex
\documentclass[conference]{IEEEtran}

\usepackage[hide]{chato-notes}

\usepackage{cite}
\usepackage{amsmath,amssymb,amsfonts}
\usepackage{algorithmic}
\usepackage{graphicx}
\usepackage{textcomp}
\usepackage{xcolor}
\usepackage{balance}

\usepackage{hyperref}
\usepackage{gensymb}
\usepackage{booktabs}
\usepackage[final]{showlabels}
\usepackage[acronym]{glossaries}

\hypersetup{
    colorlinks=true,
    linkcolor={red!10!black},
    citecolor={green!10!black},
    urlcolor={blue!10!black}
}

\newacronym{GloFAS}{GloFAS}{Global Flood Awareness System}
\newacronym{GADM}{GADM}{Database of Global Administrative Areas}
\newacronym{RF}{RF}{Random Forest}
\newacronym{GFMS}{GFMS}{Global Flood Monitoring System}
\newacronym{SMFR}{SMFR}{Social Media for Flood Risk}

\begin{document}


\title{Social Media Alerts can Improve, but not Replace Hydrological Models for Forecasting Floods}
\author{%
\IEEEauthorblockN{%
Valerio Lorini\IEEEauthorrefmark{1},
Carlos Castillo\IEEEauthorrefmark{2},
Domenico Nappo\IEEEauthorrefmark{3},
Francesco Dottori\IEEEauthorrefmark{1},
Peter Salamon\IEEEauthorrefmark{1}}
\IEEEauthorblockA{%
\IEEEauthorrefmark{1}\textit{European Commission, Joint Research Centre}%
\hspace{3em}%
\IEEEauthorrefmark{2}\textit{Universitat Pompeu Fabra}
\hspace{3em}%
\IEEEauthorrefmark{3}\textit{UniSystems SA}
\\
~\hspace{8em}Ispra, Italy \hspace{11em} Barcelona, Spain \hspace{5em} Athens, Greece \\
\{valerio.lorini, francesco.dottori, peter.salamon\}@ec.europa.eu, chato@acm.org, domenico.nappo@gmail.com
}
}

\maketitle


\begin{abstract}
Social media can be used for disaster risk reduction as a complement to traditional information sources, and the literature has suggested numerous ways to achieve this.
In the case of floods, for instance, data collection from social media can be triggered by a severe weather forecast and/or a flood prediction.
By way of contrast, in this paper we explore the possibility of having an entirely independent flood monitoring system which is based completely on social media, and which is completely self-activated.
This independence and self-activation would bring increased robustness, as the system would not depend on other mechanisms for forecasting.
We observe that social media can indeed help in the early detection of some flood events that would otherwise not be detected until later, albeit at the cost of many false positives.
Overall, our experiments suggest that social media signals should only be used to complement existing monitoring systems, and we provide various explanations to support this argument.
\end{abstract}


\input{01.introduction/introduction.tex}
\input{02.related_work/related_work.tex}
\input{03.preparing_data_labels/preparing_data_labels.tex}
\input{04.model/model.tex}
\input{05.experiment/experiment.tex}
\input{06.results/conclusions.tex}


\balance
\bibliographystyle{IEEEtran}
\bibliography{bibliography}



\end{document}

%% file: 01.introduction/introduction.tex

\section{Introduction}
\label{sec:introduction}
Even if the Paris Climate Agreement, which came into force in 2016, succeeds in keeping the global average temperature rise well below 2\degree C compared to pre-industrial levels, \textbf{global warming} is still expected to cause severe impacts.
Under the most optimistic scenario of a 1.5\degree C warming, flood damage is nevertheless set to increase by between 160\% and 240\%~\cite{dottoriincreased2018}.
%
%
%
%
Regional crisis management organizations in wealthy countries can afford the cost of high-resolution \textbf{flood monitoring systems}. On the other hand, international relief organizations with a global scope rely on global hydrometeorological models.
Different socio-economic realities combined with heterogeneous data availability (the so-called ``data divide''~\cite{gurstein2011open}) translate into various degrees of uncertainty, with reliable flood forecasts often possible only for big events.

\begin{figure}[t]
\centering\includegraphics[width=.8\linewidth]{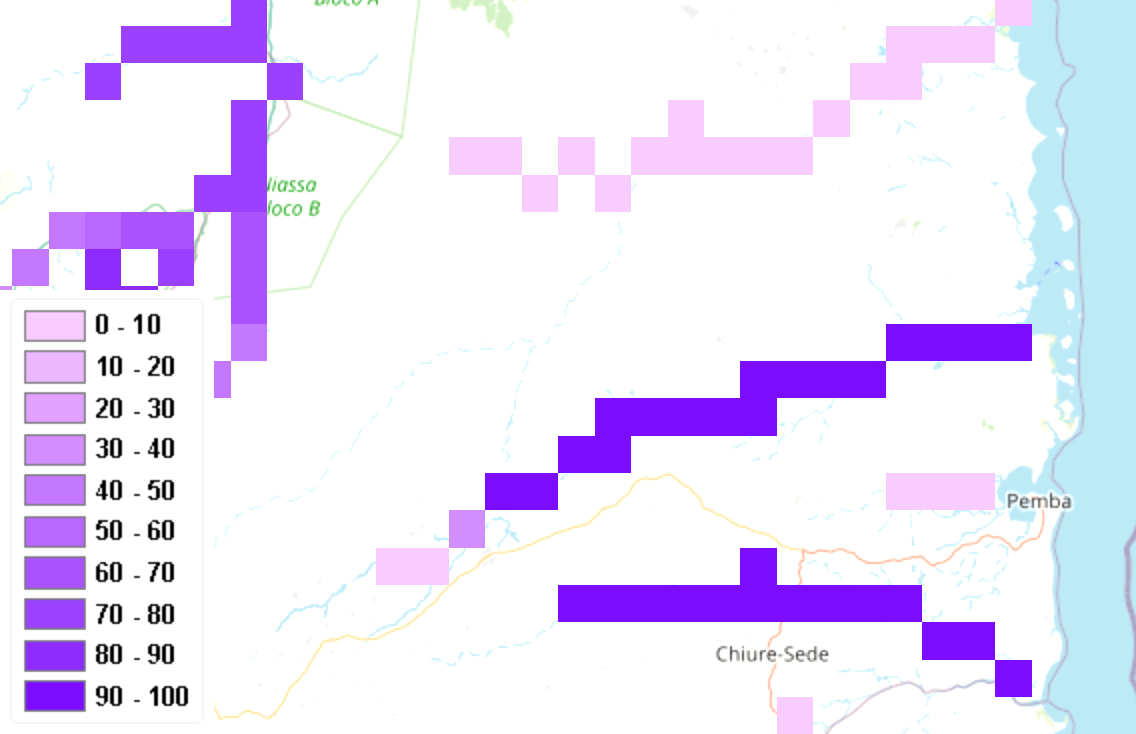}
\caption{Examples of low-uncertainty (Megaruma River in Mozambique) streamflow forecasts}
\label{strongandweak20yr}
\end{figure}

The Global Flood Awareness System (\gls{GloFAS})\footnote{\url{https://www.globalfloods.eu}} is a real-time \textbf{flood forecasting framework} that produces daily streamflow forecasts and exceedance probabilities for all major rivers worldwide\cite{PMID:32025657}.
It is part of the European Union's Copernicus Emergency Management Service,\footnote{\url{https://emergency.copernicus.eu/}} which provides information for emergency response in relation to different types of disasters.
%
%
Figure~\ref{strongandweak20yr} shows maps taken from \gls{GloFAS}, illustrating the probability of the daily streamflow forecast to exceed the local ``1 in 20-year'' discharge (i.e., the 20-year threshold, considered a severe event).
\gls{GloFAS} works by running multiple perturbed simulations, and the probability of a peak discharge exceeding the 20-year threshold is the fraction of such simulations above this threshold.
Larger values, or darker color in Figure~\ref{strongandweak20yr}, indicate a reduced uncertainty, as most simulations agree on forecasting a severe flood event for the day.

For instance, in the low-uncertainty forecast used as an example in Figure~\ref{strongandweak20yr} (top), for a few river branches almost all the simulations converge (100\% probability)
%

%
%
%
%
The uncertainty is reduced when the flow peak is predicted to occur within few days when the forecast is mostly driven by hydrological rather than meteorological conditions, and it is therefore more reliable.
High uncertainty is often associated with a large lead-time of the prediction, and the absence of a clear flood signal in meteorological forecasts.
%

Over the last decade, researchers in the field of crisis informatics have demonstrated how \textbf{social media} can be used as a relevant information source during disasters \cite{castillo2016big}.
%
%
%
This research, at the intersection of crisis informatics and disaster risk reduction, has been based largely on the extraction of public-generated discussions about flood risk in situations where weather alerts have been issued by relevant authorities, and reporting of the concerns of those impacted~~\cite{lorini2019integrating,Bruijn2020}.
Such systems are affected by the same limitations and uncertainties as the hydrometeorological forecasts themselves.

Against this background, the central question addressed by our research was:
\textbf{``Is it possible to identify floods worldwide from social media reports, using knowledge from past events and independently from hydrological forecasts?''}
Using machine learning, we created a model that takes as input the volume, trends, and characteristics of discussions about floods in social media. The output of our model is the probability that an actual flood happens, computed by supervised learning based on past events.
Because the data source of social media which we use (i.e. \emph{Twitter}), despite its large coverage and volume, produces a noisy signal that does not yield high-accuracy alerts, we cannot positively and conclusively answer the posed research question. However, our work suggests that the question may be partially answered in the affirmative, in that we can complement a flood forecasting system reducing the uncertainty of hydrological forecasts. Social Media information could be seen in this case as additional support for the Crisis Managers in the decision-making process.

In the following sections, an overview of related work is first provided, and the methods for creating a training dataset and building the model for event detection are described.
Finally, the experimental results of our work are presented, followed by conclusions and priorities for future work.

%% file: 02.related_work/related_work.tex

\section{Related work}

This Section provides an overview of some flood detection systems based on hydrometeorological models, social media or both.

\subsection{Flood detection with hydrometeorological information}

NASA's real-time Global Flood Monitoring System (\gls{GFMS}) is driven by precipitation information from the joint NASA - Japan Aerospace Exploration Agency (JAXA) satellite missions - the Tropical Rainfall Measuring Mission (TRMM) and its successor, the Global Precipitation Measurement (GPM) mission~\cite{yilmaz2010update}.
\gls{GFMS} performs rainfall analysis using a physically based hydrological model, and has a detection performance that is highest for floods of long duration and affecting a large area.
%
%

The previously mentioned \gls{GloFAS}, developed jointly by the European Commission and the European Centre for Medium-Range Weather Forecasts (ECMWF), is a global hydrological forecast and monitoring system independent of administrative and political boundaries, that is fully operational within the EU’s Copernicus Emergency Management Service.
GloFAS couples weather forecasts with a hydrological model to produce daily flood forecasts.
Due to its meteorological forcing (i.e., rainfall map, wind speed map, temperature map, etc.) and spatial resolution of 0.1 degrees, GloFAS performs well for large rivers. The lack of finely distributed meteorological observations at a global scale limit the resolution of the calibration for the forecasting for smaller rivers~\cite{HIRPA2018595}.

%

\subsection{Flood detection from social media}

In recent years research on the use of citizen-generated information to provide near real-time information during crisis situations, has received increasing attention.
In particular, the public stream of postings from Twitter has created an opportunity for crisis responders to ingest crisis-relevant messages into reports during natural hazards~\cite{olteanu2015expect}.

The study of social media during floods has unique challenges compared with its use for other types of disasters such as earthquakes.
For the latter, there are records from seismographic monitoring networks using widely accepted standards, and therefore reliable lists of georeferenced events are available~\cite{robinson2013sensitive,poblete2018robust}.
For flood events, however, there is no such international standard for recording and reporting information, let alone any unique identification number of these events. Instead, emergency managers report information according to their own interpretation and local guidelines.
Despite these limitations, previous work has demonstrated how social media can be used to detect floods~\cite{brouwer_probabilistic_2017}, with the aim of augmenting situational awareness~\cite{ROSSI2018145}.%

%
%

%
%
%
We use a validated reference set of events tracked by independent organizations, covering a wide range of events in terms of geographical region, duration, extent and magnitude.

Based on this we have built a supervised model for catching signals from social media on heterogeneous types of flood (riverine floods, coastal floods, flash floods, hurricane floods, etc.) at a sub-national scale.
We then analyzed our results in the light of forecast information available at the time, just before the event, in order to understand if social media information can represent an added value to those systems.
%

%% file: 03.preparing_data_labels/preparing_data_labels.tex

\section{Dataset preparation}
\label{sec:preparing_data}

The ground truth data that we used is constructed at the level of days and countries, with each record indicating if there was a confirmed flood on that day in that country.
We collected social media data and then processed it to the same level of granularity by extracting various features. Further details on the dataset preparation are presented below.

\subsection{Ground Truth}
\label{subsec:ground_truth}

Because no single comprehensive database exists that containing all worldwide floods, we used a list of flood events from previous research that aggregates data from different sources.
For details on this list of events, the interested reader can consult the original source~\cite{lorini2020uneven}.
Briefly, the list of flood events is collected from three different databases: Europe's Floodlist; the UN's Emergency Events Database (EM-DAT), and the Dartmouth Flood Observatory (DFO) of the University of California.
%
%

\begin{figure}[t]
\centering\includegraphics[width=.9\linewidth]{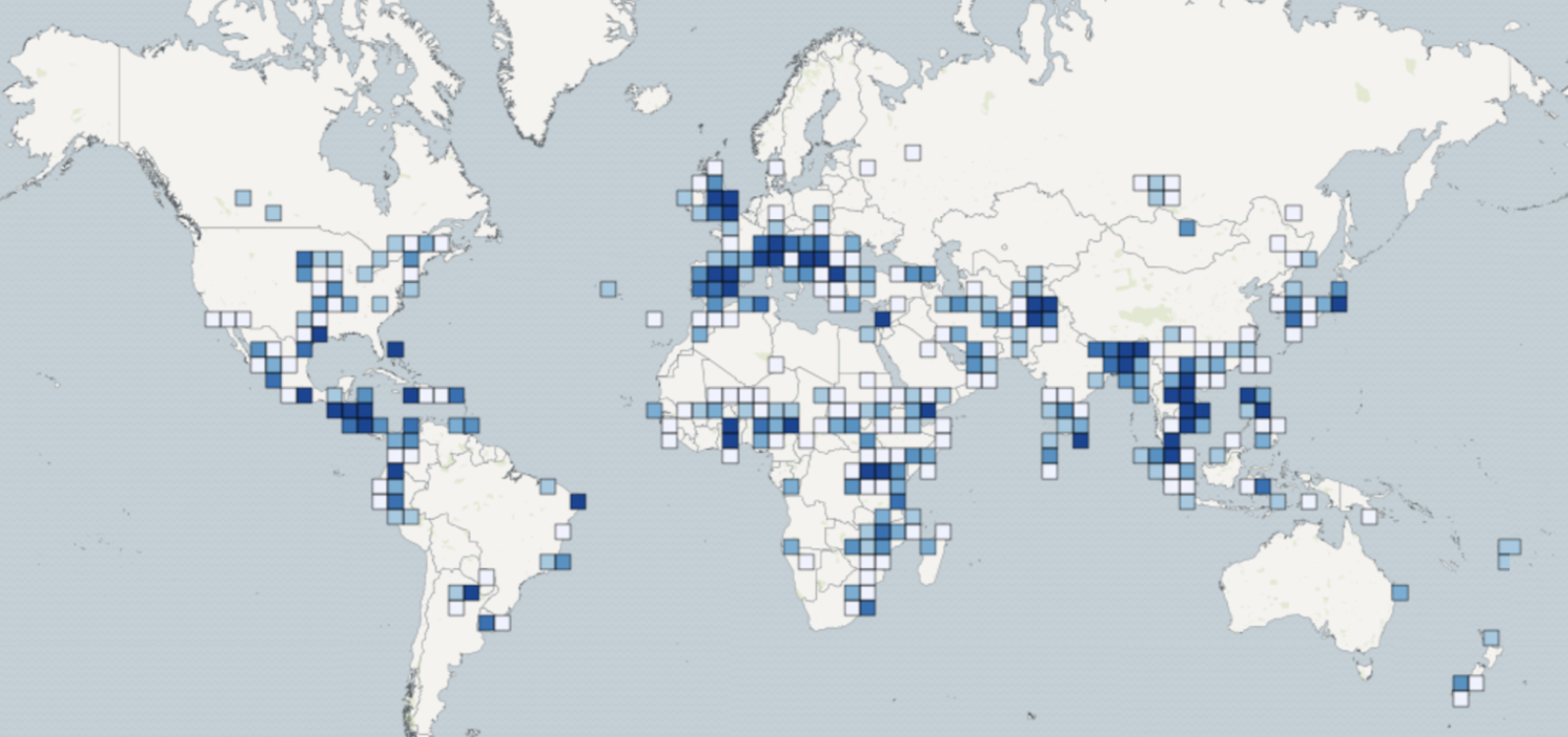}
\caption{Spatial distribution of the flood events used as ground truth. Darker color indicates multiple events in the same administrative area}
\label{fig:mapevents}
\end{figure}

We selected all events that could be geocoded to a sub-national level. This led to 349 events spread over 1,318 administrative areas, as shown in Figure~\ref{fig:mapevents}.

\subsection{Data Collection}
\label{subsec:collection}

We collected tweets relying on the public Twitter streamer.\footnote{\url{https://developer.twitter.com/en/docs/tweets/sample-realtime/overview/get_statuses_sample}}
We opted to collect posts on floods using a set of flood-related keywords in several languages (i.e., English, Spanish, Italian, German, French, Portuguese, Arabic) for a nine-month period, covering flood-seasons worldwide.
The complete list of keywords is available in our data release.
Our data collection period was April to December 2019, but was interrupted frequently.
We experienced network, software, and hardware failures, together with limitations applied by the streamer provider.
We worked around the multiple interruptions during the 274 days of observation, by applying the following heuristic method:
if for any given day there were more than six hours without any tweet, we would mark that entire day as ``invalid''.
In our observations, this was always an indication of some kind of failure.
%
The days for which we have data were 74\% of the total observation days.
%

\subsection{Features}
\label{subsec:features}
Each collected tweet was automatically annotated as either flood-related or not flood-related by a multi-lingual classifier~\cite{lorini2019integrating}.
The flood-related tweets were then geocoded using their available geographical metadata, and when this was not present (i.e. in the majority of cases), using place-names mentioned in the text\cite{halterman2017mordecai}.
Next, we aggregated individual tweets in space (i.e. the affected area) using the Database of Global Administrative Areas (GADM) spatial database of the location of the world's administrative areas.\footnote{\url{https://gadm.org}}

We aggregated tweets according to days, to match the granularity of our ground truth data.
\begin{table}[ht]
\label{tab:multicol}
\centering
\caption{Features extracted; $p_i$ is the probability that tweet $i$ is related to floods, as computed by an automated classifier.}
  \begin{tabular}{p{.3in}p{2.9in}}

    \toprule
    \multicolumn{2}{c}{\textbf{Keys}}\\
    \midrule
    Name & Description\\
    \midrule
Date $d$ &
  Day number \\
Region &
  Administrative region in GADM \\
  \midrule
    \multicolumn{2}{c}{\textbf{Daily features (22 feat)}}\\
    \midrule
Lang &
  0: English not an official lang, 1 or 2: English is 1st or 2nd lang \\
TOT &
  Tot number of Tweets on this day and this region \\
T00 &
  In bucket T$a$-$b$, number of postings having $a < p_i \le b$ \\
P00 &
  In bucket P$a$-$b$, fraction of postings having $a < p_i \le b$ \\
    \midrule
    \multicolumn{2}{c}{\textbf{Lagged features (50 feat) computed over a moving window of 3 days}}\\
    \midrule
T3P00 &
  In bucket T3P$a$-$b$, total number of postings having $a < p_i \le b$ on days $\{ d, d-1, d-2 \}$ \\
M3P00 &
  In bucket M3P$a$-$b$, fraction of postings having $a < p_i \le b$ on days $\{d, d-1, d-2 \}$; these add up to 1.0 \\
A3P00 &
  In bucket A3P$a$-$b$, average fraction of postings having $a < p_i \le b$ computed over the three days \\ 
D1T00 &
  In bucket D1T$a$-$b$, change in the number of postings having $a < p_i \le b$ between day $d$ and day $d-1$ \\
I3T00 &
  In bucket I3T$a$-$b$, maximum increase in the number of postings having $a < p_i \le b$ between day $d$ and day $d-1$, or day $d-1$ and day $d-2$; this is always non-negative \\
    \bottomrule
\end{tabular}
\end{table}

For each day and region, we measured the features listed in Table~\ref{tab:multicol}.
%

In order to enable comparisons between regions with different population sizes and different degrees of Twitter adoption, we produced normalized features by considering the average number of postings originating from each region during a period of one month.
We calculated such values by analyzing one month of geolocated data extracted from the Public Streamer available on the Internet Archive digital library.\footnote{\url{https://archive.org/}}
All of the 72 features in Table~\ref{tab:multicol} were divided by the expected number of postings for the same region, with the exception of features P00-10 $\dots$ P90-100 and T3P00-10 $\dots$ T3P90-100, which reflect proportions, and the ``language'' feature.
In addition, the expected number of postings for the region was added to the feature list, for a total of 73 features.

\subsection{Data Labeling}

In order to create the training data, it is not sufficient merely to associate each row in the dataset table (corresponding to a date and a region), with a label of flood (\emph{True}) or no-flood (\emph{False}).
The main reasons for this are that flood events and the related discussion on social media generally last more than one day, and they build up over time.
Furthermore, there is no common, widely accepted definition of when a flood starts or ends.
Indeed, in the original data sources used as ground truth, when two or more sources have the same event, the dates are not necessarily the same.
%
%
Some ambiguity is inevitable at testing time, but in the training data, we would like to learn only from unambiguous cases.

\begin{figure}[t]
\centering\includegraphics[width=.8\columnwidth]{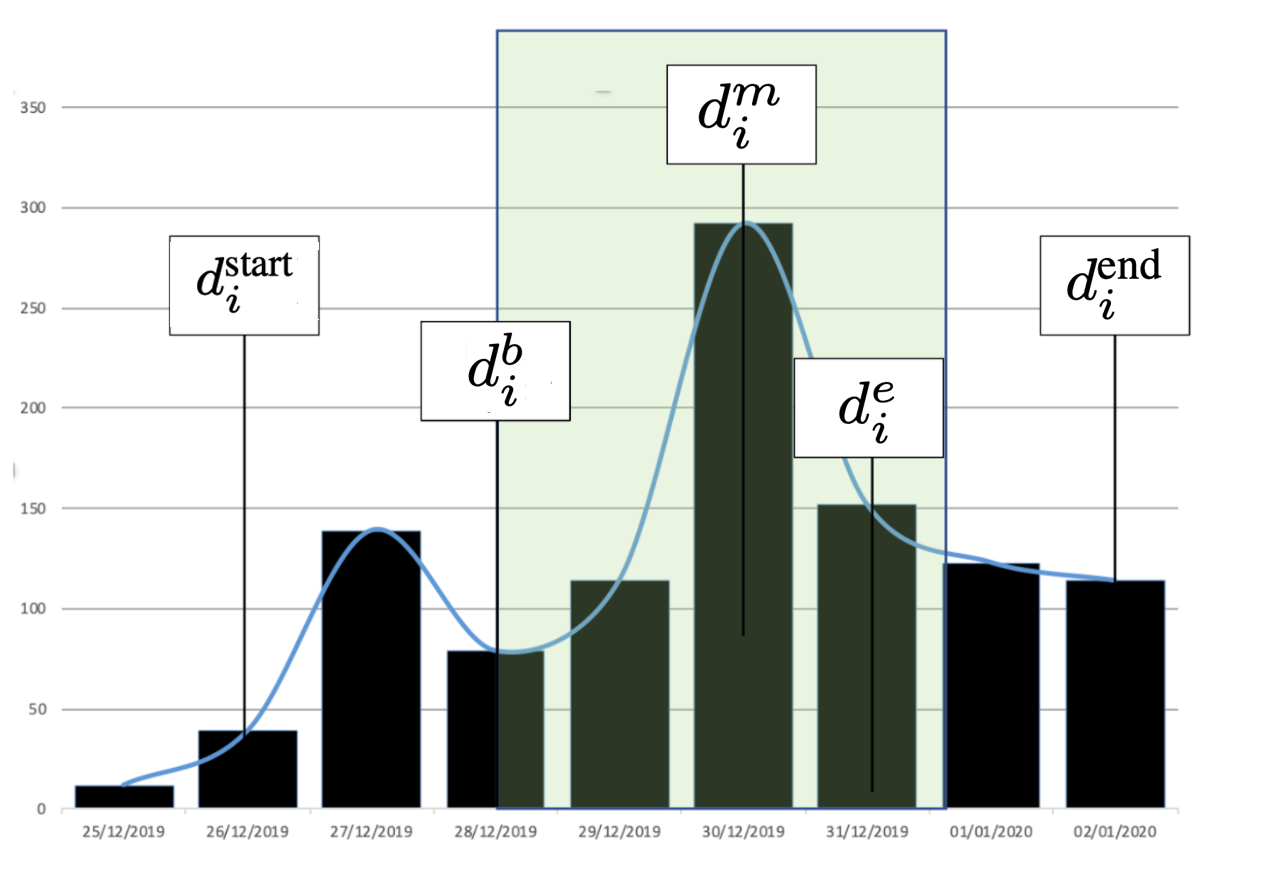}
\caption{Overall volume of flood-related postings per days overlapping with a flood event i lasting eight days from ``Event Start'' ($d_i^{\text{start}}$) to ``Event End'' ($d_i^{\text{end}}$).
The range that is labeled as \emph{True} in the training data goes from $d_i^{b}$ to $d_i^{e}$ and reaches its peak at $d_i^{m}$ (shaded area)}.
\label{fig:peakdate}
\end{figure}

Our methodology associates the floods in our ground truth with specific time-spans (days) and regions.
Then, we consider the dynamics of the discussion following the evolution of an event $i$, spanning between dates $d_i^{\text{start}}$ and $d_i^{\text{end}}$ in the ground truth.
Figure~\ref{fig:peakdate} shows sample data for one such flood event.
In Figure~\ref{fig:peakdate}, the overall number of social media postings in the same region is represented by bars.
To associate labels with days and regions in the training data, we consider a range of days within the series corresponding to the same region in which a flood was recorded in the ground truth.
This range is created as follows:
\begin{enumerate}
\item We locate the day with the local maximum or peak $d_i^{(m)} \in [d_i^{\text{start}}, d_i^{\text{end}}]$ of social media activity within the days of the flood according to the ground truth.
\item We locate the beginning of that increase in activity, i.e $d_i^{(b)}$.
\item We set the end of the range to be the day after the maximum, $d_i^{(e)} = d_i^{(m)}+1$, because in our observations there is almost invariably some conversation about the flood that remains in social media after the peak.
\end{enumerate}

Then, we set all days within that region lying in the range $[d_i^{(b)}, d_i^{(e)}]$ to \emph{True}.

A period of 10 days before and 10 days after the ground truth for the flood in the same administrative region is a grey area, due to the fact that increases social media in activity may or may not be present.
Hence, we set all days within $[d_i^{\text{start}}-10, d_i^{\text{start}}-1]$ and $[d_i^{\text{end}}+1, d_i^{\text{end}}+10]$ to \emph{Undefined}.
The remainder of the days in this region are set to \emph{False}.

We also have to account for ambiguities in geocoding, which may associate floods in one administrative region with another administrative region in the same country.
To preclude these from occurring in our training data, we remove labels from all other regions of a country where no floods are recorded in this period, in the period $[d_i^{\text{start}}-5, d_i^{\text{end}}+20]$.
%
%
For regions where no floods are recorded within the entire observation period, we set all labels for all days to \emph{Undefined}.

Finally, we considered only regions where English was an official language, and only days in which the total number of collected tweets was above 100.
Our final training data contains 930 \emph{True} or \emph{False} rows corresponding to (day, region) pairs, of which 73 (or 7.9\%) have the label \emph{True}.
%

%% file: 04.model/model.tex

\section{Building the Model}
\label{sec:building}

We posed our research question as a binary classification problem, in other words detecting from the features extracted from postings whether these corresponded to a day and region with floods or without floods.
We experimented with various classification schemes including Support Vector Machines, Multi-Layer Perceptron, and Random Forests.
Random Forest (RF) classifiers yielded the best results.

We performed a grid search to optimize the learning parameters.
For feature selection, we used an ANOVA F-Test, with the best performance obtained by selecting 40 out of the 73 features.
The selected features using univariate feature selection with a classification function score covers most of the features classes in Table~\ref{tab:multicol}.
They describe a combination of aggregated classes and average probability: 'T00-10', 'T10-20', 'T20-30', 'T30-40', 'T40-50', 'T60-70', 'T70-80', 'T80-90', 'T90-100', 'P00-10', 'M3P00-10', 'M3P90-100', 'A3P00-10', 'A3P60-70', 'A3P90-100', 'T3P00-10', 'T3P10-20', 'T3P20-30', 'T3P30-40', 'T3P40-50', 'T3P50-60', 'T3P60-70', 'T3P70-80', 'T3P80-90', 'T3P90-100', 'D1T10-20', 'D1T20-30', 'D1T30-40', 'D1T40-50', 'D1T70-80', 'D1T80-90', 'D1T90-100', 'I3T00-10', 'I3T10-20', 'I3T20-30', 'I3T30-40', 'I3T40-50', 'I3T70-80', 'I3T90-100', 'TOT'

For the \gls{RF} parameters, we obtained the best results with 1,000 decision trees and a maximum depth of two levels for each tree, although we observed that similar numbers of trees and depths did not yield a substantively different performance.
We used three-fold cross-validation using two-thirds of the data for training and the remaining one-third for testing in each iteration.

\begin{figure}[htb!]
\centering\includegraphics[width=.9\columnwidth]{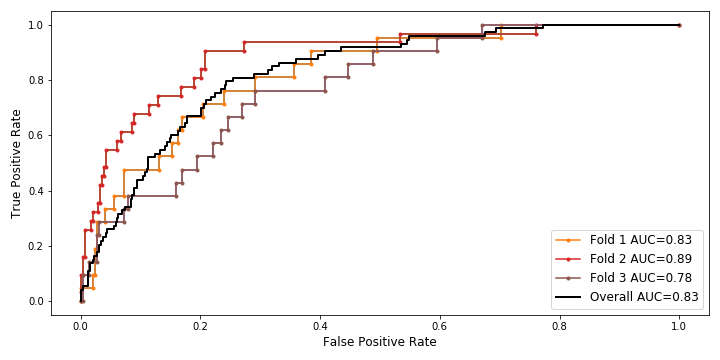}
\caption{ROC curve of the obtained classifier.}
\label{fig:runsmodels}
\end{figure}
Figure~\ref{fig:runsmodels} shows the Receiver Operating Characteristic (ROC) curve representing the tradeoff between sensitivity (i.e. the true positive rate, or how good the model is at detecting real floods) and specificity (i.e. the true negative rate, or the model’s ability of avoiding false alarms).
An increase in sensitivity is accompanied by a decrease in specificity. We observed that the best model does not offer a high-sensitivity high-specificity classification.
By setting a threshold for a positive (i.e. ``flood'') classification of 0.2, which yield a good balance of precision and recall, we obtain an average combination of precision of 34\% and recall of 41\%.
This implies that this model cannot be used independently of hydrological flood monitoring systems for detecting floods.
However, as we will show in the following Section, the \gls{GloFAS} forecasting performance analyzed during the ground truth events would benefit from our classifier in terms of reducing uncertainty.

%% file: 05.experiment/experiment.tex

\section{Leave-One-Out Experiment}
\label{sec:zero-shot}
In the previous Section~\ref{sec:building} we calculated the Accuracy of our model using statistic.
The model presented, as described in Section~\ref{sec:introduction}, aims at an operational use together with global forecasting systems.
In order to measure the accuracy of our model in an hypothetical operational system, we need to define the hit-rate as the percentage of real events that our model could predict.
We simulated such methodology using a Test/Train split supported by group definition. In our previous experiments we considered the rows labeled as '1' uncorrelated from each other. Although this is true, relying solely on random split could lead to bias test data towards a specific event (multiple rows).
In other words, for this experiment we assigned an 'event\_id' to each of the rows of the dataset and we grouped and isolated each time an event to define the Test dataset as this would be the case for future events.
Hence, we consider how this model would perform in a ``leave-one-out'' cross-validation scenario, and particularly, whether it can complement the forecasts of \gls{GloFAS}.

\subsection{Overall Results}

We first observed that the rows (days and regions) labeled \emph{True} are correlated with each other, if they occur in the same region around the same time, as in the case that they represent the same flood.
Hence, we cannot leave out one row, but instead must leave out an entire event.

Since we wish to perform a side-by-side comparison against an operational system for disaster risk reduction, we consider two possible outcomes: ``Hit'' or ``Miss''.
The former is when we trigger an alert for at least one of the days of a flood, while the latter is when we do not.

\begin{table}[ht]
\caption{Output of the leave-one-out classifier and \gls{GloFAS} for 23 flood events.}
\label{tab:leaveoneout}
\centering
\resizebox{.5\textwidth}{!}{%
\begin{tabular}{llrccl}
    \toprule
    Place&Country&Days&Result&GloFAS 20yr&Type of event\\
    %
    %
    %
    \midrule
    Suffolk&USA&5&miss&no river&Storm surge\\
    Herkimer&USA&2&miss&10-20\%&Heavy rain, flash floods\\
    %
    %
    Hartbeespoort Dam&SouthAfrica&3&hit&30-40\%&Heavy rain, flash floods\\
    \bottomrule
\end{tabular}
}%
\end{table}

The hit rate of the experiment for the 23 simulations done (i.e. one per each event in our training data) is 52\%, which means that we capture about half of the flood events.
Table~\ref{tab:leaveoneout} shows three cases from October and December 2019.
%


On a cautionary note, it is important to bear in mind that the main purpose of \gls{GloFAS} is to forecast riverine floods, and therefore those events that are a combination of riverine, coastal and / or flash floods can only be compared to \gls{GloFAS} forecasts to a limited extent.
%
%
%

%
Our methodology offer the obvious advantage to capture all types of flooding (coastal, flash flood, pluvial, etc.) with the same ML-trained model.
We observed that in many of the cases our model indicates flood activity, although this must be considered in light of the computed average precision of 34\% (described in the previous section), meaning that about one in three of the alarms generated by the model based on social media alone will correspond to a flood.

\subsection{Case Studies}

In order to better understand the hit-rate performances of the leave-one-out model, we analyzed some of the cases in detail.

Firstly, the floods of October 2019 in \textbf{Suffolk county} in the US were missed by our model:\footnote{\url{https://bit.ly/2UoOxg9}} ``Minor flooding was reported in parts of Suffolk County, New York. Roads were swamped and some buildings flooded''.
In this case, the flooding can be considered minor, as neither fatalities nor injured persons were recorded in the ground-truth dataset. Our system would have missed this flood, and \gls{GloFAS} indicated no signal.

Secondly, the floods of October 2019 near \textbf{Herkimer (Mohawk Valley)} in the US were also missed by our model:\footnote{\url{http://floodlist.com/america/usa/halloween-storm-flood-october-2019-new-york}} ``According to a statement by New York Governor Andrew Cuomo’s office on 01 November, over 240,000 homes were without electricity and nearly 60 roads were closed''.
Although this event affected more people, the main impact was an electricity blackout. Our system would have also missed this flood, and \gls{GloFAS} indicated a 10-20\% chance of exceeding the 1 in 20-year discharge.
At the peak of the storm, we found that 75\% of the total postings were classified as not relevant to floods resulting in low values for features used by our model.

In both of these cases, the reason for which the \gls{GloFAS} model had little or no signal was that the main driver of the flood was water from the storm and from storm surge (coastal flood), rather than water overflowing from a river.


Thirdly, the floods of December 2019 in \textbf{Gauteng and North West Provinces (near Hartbeespoort Dam)} in South Africa were captured by our model:\footnote{\url{http://floodlist.com/africa/south-africa-floods-gauteng-december-2019}} ``Hundreds or people have evacuated their homes. News 24 reported that one person died on 9 December when flash floods swept a vehicle from a low-lying bridge close to Hartbeespoort Dam in North West Province, about 35 km west of Pretoria.''.
In this case, \gls{GloFAS} forecasted a probability of an extreme (1 in 20-year) event in the area, of 30-40\%. Our test data features indicate more than 2,500 social media postings in one day, 40\% of which were classified as highly relevant.
%

%
%

%% file: 06.results/conclusions.tex

\section{Conclusions}

While the forecasting of floods using hydrometeorological models is possible within certain limits, many floods are not forecasted or are forecasted with only a low probability.
Although comparing forecasts from a hydrological model would be fairer if  only purely riverine floods were considered, only in one case of the flood events taken into consideration the computed probability of exceeding a '1 in 20-year' threshold forecast by \gls{GloFAS}, was more than 50\%.
This is largely because forecasting systems are based on model simulations, meaning that they are affected by noisy signals due to many factors (e.g., noise in meteorological forecasts, missing data, and incomplete reference data).

\begin{figure}[h]
\centering\includegraphics[width=.7\columnwidth]{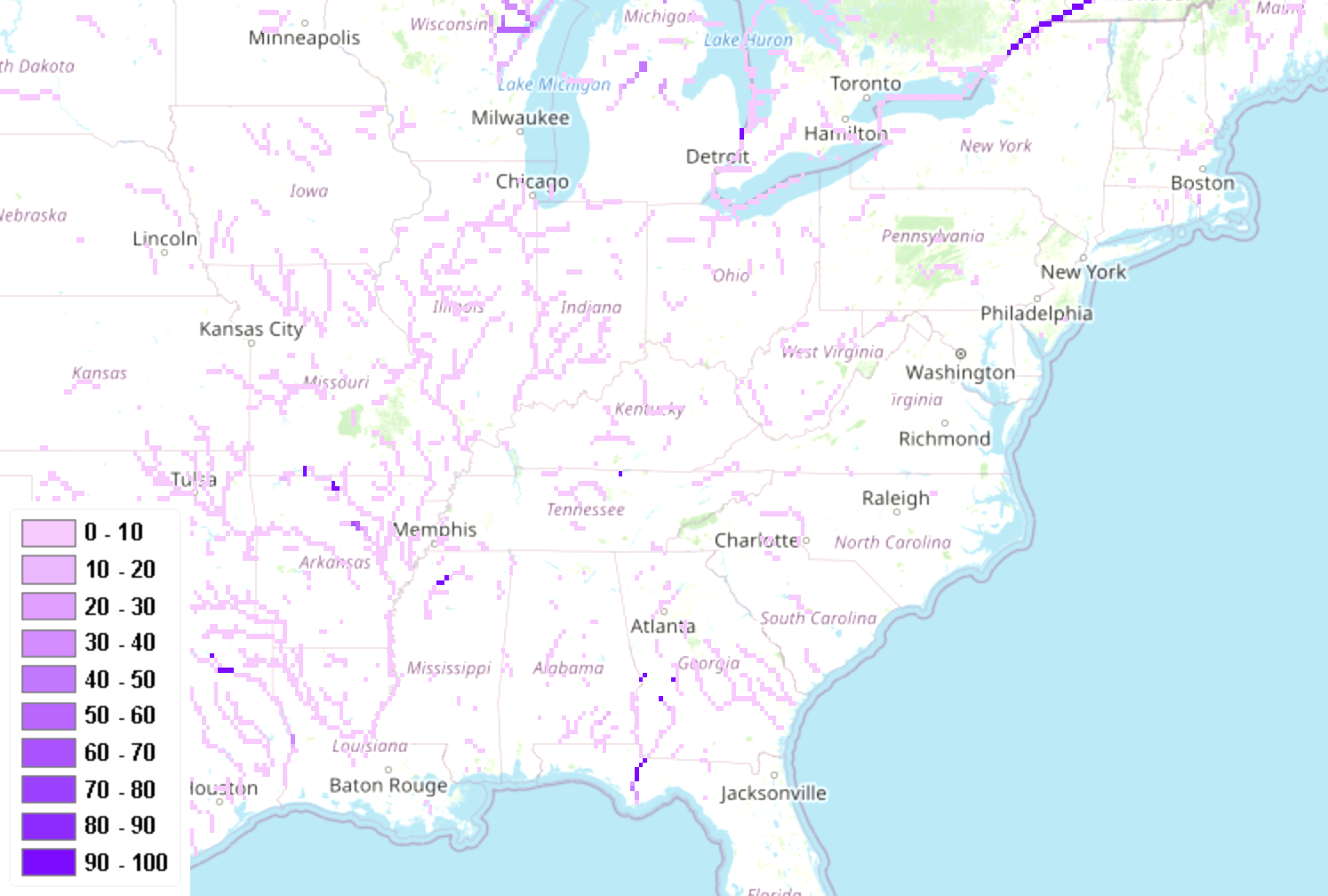}
\caption{\gls{GloFAS} map for April 2020, highlighting in purple portions of river basins that have a heightened probability of floods according to darkness: this happens in many areas in the US at the same time. (Better seen in color)}
\label{fig:mapapril}
\end{figure}

Both forecasters and emergency managers require tools that can help them to narrow down uncertainty.
For example, Figure~\ref{fig:mapapril} shows the ensemble probability of exceeding the 1 in 20-year flood threshold in the US for April 2020, as forecast by \gls{GloFAS}.
As can be seen, \gls{GloFAS} forecasts low to medium probability of a flood in many administrative areas.
Given the importance of making well-informed decisions, our research has partially affirmed our initial question (i.e., the possibility to detect floods worldwide from social media reports), since the results of the leave-one-out experiment indicate that \textbf{our model can indeed spot impactful events where damages are clearly related to water}.

The methodology proposed for leveraging our model in real-time is to keep the collection of tweets as described in section ~\ref{subsec:collection} running in the background and tasking the model classification on a daily basis.
Since the model uses features with data from the previous two days,  we think it can  ’detect’  a  change in the conversation on social media as shown in Figure~~\ref{fig:peakdate}.
The slope of the peak can be smoother or steeper according to the type of event,  the classifier could be able to detect the event before its peak,  which is associated with the highest impact.
Our work provides an added value to the GloFAS hydrometeorological forecasts since it helps to reduce uncertainty and broadens the range of flood-types that can be detected

In particular, we are confident that the precision of our classifier in determining whether a flood is occurring in a specific area, could be improved by: (a) using our model as a trigger for a more focused real-time data collection, where city names are used instead of flood-related keywords, and (b) setting threshold levels for the ratio between tweets classified as ``most likely flood-related'' and those classified as ``likely flood-related'', for a real time aggregation of data. In the latter case, the ratio can act as an indicator to filter out noise caused by trending topics in the specific area (i.e., sport events, celebrities, politics, etc.), where we expect to have more tweets unlikely to be about floods.

Another potential improvement to be addressed in future work, concerns the normalization of features. In our study, we normalized the features of the training dataset using the number of expected data at the national level. However the adoption of a particular social media platform may vary between regions within the same country. Another practical issue that merits further investigation is how to handle multiple crisis events (not necessarily all natural hazards) happening in the same country. When there is a strong trending topic, we observed that floods might receive less attention from mass media and from those members of the public who are not directly affected, which reduces the strength of the social media signal.

Our research has demonstrated that the methodology proposed can complement a global flood forecasting system. One aspect which we have not yet addressed is the potential link and research related to the concept of lead time of the forecast. In other words, so far we have analyzed the additional value of the model on the day of the event, while there is still potential improvement in considering the forecast of an event, specifically when it builds over several days.
\paragraph*{\textbf{Reproducibility}}
All of our code, as well as data to reproduce the results on this paper, are  available for research purposes with the camera-ready version of this paper\footnote{\url{https://zenodo.org/record/4274495}}.

\paragraph*{\textbf{Acknowledgments}}
Castillo thanks La Caixa project LCF/PR/PR16/11110009 for partial support. The authors thank Javier Rando for his help on collecting data for the ground truth, and Niall Mccormick for proofreading this manuscript